# Correlated Electrons Step-by-Step: Itinerant-to-Localized Transition of Fe Impurities in Free-Electron Metal Hosts

C. Carbone[1], M. Veronese[1], P. Moras[1], S. Gardonio[1], C. Grazioli[1], P. H. Zhou[2], O. Rader[3], A. Varykhalov[3], P. Gambardella[4], S. Lebèque[5], O. Eriksson[6], M. I. Katsnelson[7], A. I. Lichtenstein[8]

[1]Istituto di Struttura della Materia, Consiglio Nazionale delle Ricerche, I-34012 Trieste, Italy
[2]International Center for Theoretical Physics, I-34014 Trieste, Italy
[3]BESSY GmbH, D-12489 Berlin, Germany
[4]ICREA and Centre d'Investigacions en Nanociència i Nanotecnologia (CIN2-ICN), E-08193 Barcelona, Spain
[5]Laboratoire de Cristallographie, Rèsonance Magnètique et Modèlisations, Institut Jean Barriol, Nancy Universitè, 54506 Vandoeuvre-lès-Nancy, France
[6]Department of Physics and Materials Science, Uppsala University, Box 530, 75121 Uppsala, Sweden
[7]Institute for Molecules and Materials, Radboud University of Nijmegen, 6525AJ Nijmegen, The Netherlands
[8]Institute for Theoretical Physics, University of Hamburg, Germany

High-resolution photoemission spectroscopy and realistic *ab-initio* calculations have been employed to analyze the onset and progression of *d-sp* hybridization in Fe impurities deposited on alkali metal films. The interplay between delocalization, mediated by the free-electron environment, and Coulomb interaction among *d*-electrons gives rise to complex electronic configurations. The multiplet structure of a single Fe atom evolves and gradually dissolves into a quasiparticle peak near the Fermi level with increasing the host electron density. The effective multi-orbital impurity problem within the exact diagonalization scheme describes the whole range of hybridizations.

Electronic states in solids exhibit either itinerant or localized behavior depending on a number of factors, such as type and strength of the chemical bonding, local atomic arrangement and dimensionality of the system. This duality represents a long-standing challenge in solid state physics, barring a general theory of electron transport and magnetic phenomena, as well as in other areas of science including astrophysics [1]. The standard band theory based on the Bloch theorem [2] describes very well many properties of metals and semiconductors which can be generally explained by Landau's Fermi-liquid theory with its postulate of a one-to-one correspondence between states of bare particles and quasiparticles [3]. At the same time, electron states in atoms are classified in terms of *many-particle* quantum numbers of total spin ($S$), orbital ($L$) and total angular ($J$) momentum forming a multiplet structure. It is intuitively clear that in some cases this description should survive in solids as well, assuming that an overlap of atomic states is small enough. There is, however, no simple way of relating the many-electron atomic states with the single-particle Bloch waves.

The established theory of atomic multiplet structure, which started from the seminal works of Racah [4], describes with great accuracy the complex spectra of transition-metal ions [5]. The main mechanism of energy level formation in this case is related to the strong Coulomb interaction among 3$d$-electrons. The factor determining whether an atomic multiplet structure should form, or if energy bands are to be expected, is the competition between the Coulomb energy and the kinetic energy associated with electrons hopping from site to site in the lattice. When neither of these two terms dominates, one often observes complicated electron states that are manifestations of electron correlation. Electronic correlation is a crucial ingredient not only in the narrow $d$-band systems, but also in rare-earth [6], actinide [7], and even some *sp*-electron compounds, like the superconducting fullerene family [8].

Narrow quasi-localized $d$-states play an important role in the electronic structure of low-dimensional transition-metal systems, including surfaces and nanoparticles [9]. The most prominent many-body effect, the Kondo resonance, related with single transition-metal impurities on the surface of simple metals, can be directly visualized by scanning tunneling spectroscopy [10, 11]. When nanosized materials approach the atomic limit one may hence expect that conventional theoretical models fail, and the question is how to describe these systems appropriately. For instance, the magnetic moments of single Co atoms on a Pt surface cannot be reproduced using conventional theoretical methods, which ignore electron correlations [12]. Indeed, the case of individual magnetic impurities is highly nontrivial and may be considered as a benchmark for understanding electron correlation in $d$-metal systems. For moderate hybridization between impurity and host electron states the Anderson model describes well the formation of a magnetic moment and the many-body spin-flip processes that lead to the Kondo effect [13]. However, for rare-earth impurities one should start with the multiplet structure of a free atom and then introduce hybridization effects as a perturbation [14]. These two approaches are closely related with the limiting cases of "strong" and "weak" hybridization, whereas a generic intermediate case, schematized in Fig. 1, seems to be very difficult to understand.

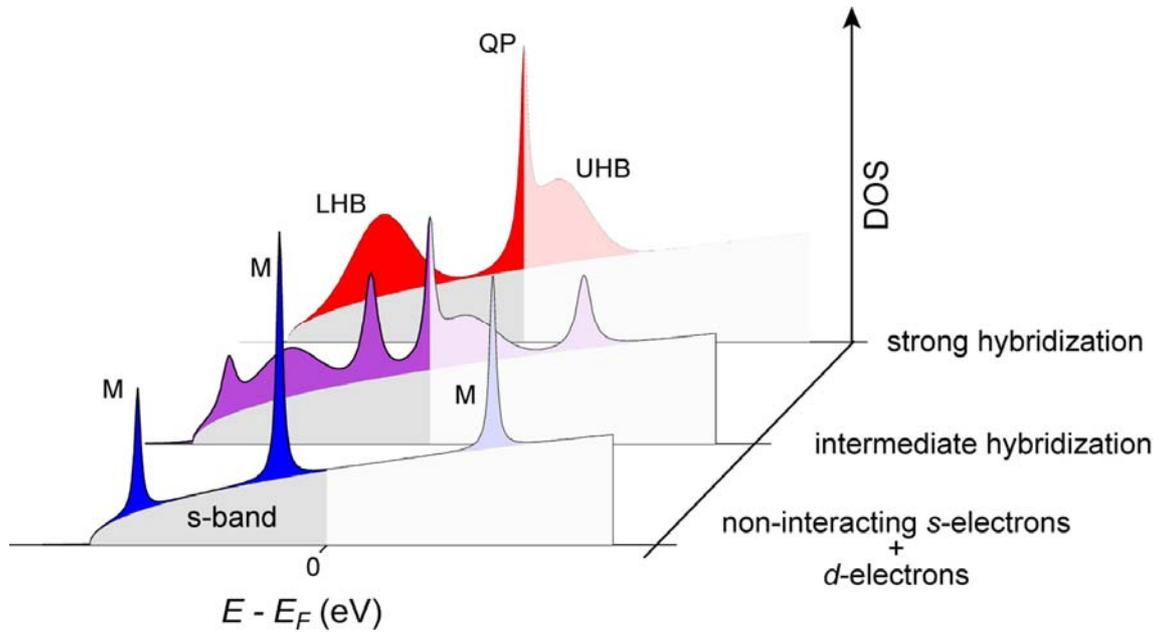

**Fig**. **1**. Spectral function of a many-body *d*-electron impurity and host conduction electrons. Depending on the local Coulomb interaction and hybridization strength different behaviors are expected: localized impurity limit (bottom); intermediate coupling (middle); strong hybridization limit (top). Letters indicate multiplet (M), quasiparticle resonance (QP), lower and upper Hubbard bands (LHB, UHB) spectral features.

Direct spectroscopic methods can be used to probe in detail the transition from an atomic-like multiplet structure to a band energy spectrum. Nevertheless, no experimental or theoretical study has shown how the transition occurs between atomic-like and itinerant electron configurations, when the relative importance between electron hopping and Coulomb repulsion is carefully tuned. The experimental verification of multiplet-to-band transitions is a delicate task that requires appropriate choice of a *d*-metal and conduction electron "bath" systems. Here we focus on Fe impurities bound to different alkali surfaces, which represent a realistic approximation to a free-electron metal and whose charge density can be varied step-wise by moving along the alkali group in the periodic table [2].

Experiments were performed at the ID4 and PGM 56 beamlines of the Elettra (Trieste) and BESSY (Berlin) synchrotron radiation facilities. Photoemission electron spectroscopy was employed to measure the valence band electron configuration of dilute Fe impurities on multilayer alkali metal films quench-condensed on single-crystal Cu(100) in ultra-high vacuum (base pressure $< 2 \times 10^{-10}$ mbar). Photoemission spectra were recorded using 50-120 eV photon energy in normal emission with 8º angular acceptance and 15 meV energy resolution. In order to avoid aggregation, minute amounts of Fe atoms were deposited and measured at 20 K, below the onset of thermal diffusion. The Fe surface concentration is indicated in the text as a fraction of a monolayer (1 ML = $1.6 \cdot 10^{15}$ atoms cm$^{-2}$).

Fig. 2 shows the photoemission spectra of dilute Fe impurities on a K film as a function of Fe coverage recorded with photon energy hν = 120 eV. In our experimental conditions, the spectral features of the Fe 3$d$-states can be clearly detected even at very low atomic concentrations owing to the favorable cross-section ratio with respect to the alkali $sp$-states [15]. The Fe spectra display clear signatures of multiplet structures around -3 and -0.3 eV below the Fermi level ($E_F$), whose lineshape is unchanged up to the impurity coalescence threshold of about 0.035 ML. In agreement with a core level X-ray absorption study of Fe impurities on K [16], these features are assigned to a nearly atomic-like Fe $d^7$ configuration, representing strongly localized high spin and low spin terms superposed to the $sp$ substrate bands.

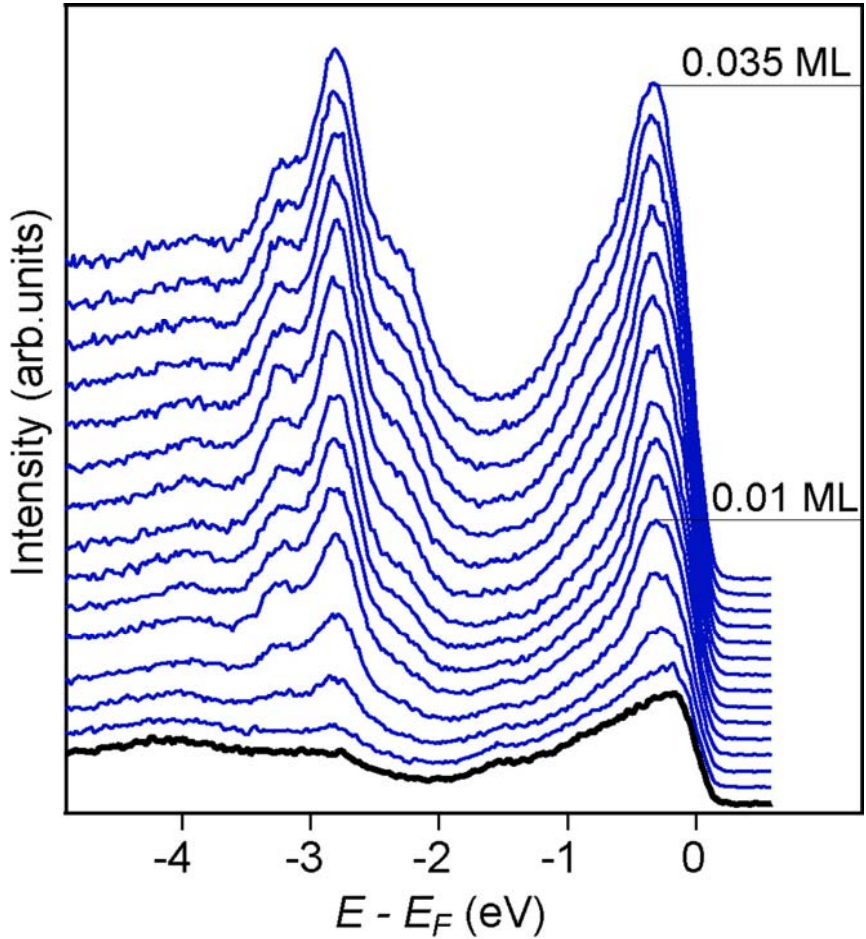

**Fig**. 2. Photoemission spectra of Fe impurities on a K film. The bottom spectrum represents the substrate photoemission background. The intensity of the Fe-induced spectral features increases with Fe coverage from bottom to top. No changes in the spectral lineshape are observed in the isolated impurity limit below 0.035 ML.

Depending on the atomic volume of the alkali ions, the surface electron density can be decreased or increased by moving towards heavier and lighter alkali species, respectively. We find the hybridization of the Fe *d*-states to change drastically from being very weak for Cs to much stronger for Li. Fig. 3 shows photoemission spectra for 0.01 ML Fe atoms on Cs, K, Na, and Li hosts after subtraction of the surface background. From Cs to K only a reduction of intensity of the multiplet features is observed accompanied by a moderate energy broadening. On Na and Li, however, the spectra change *qualitatively*, indicating the onset and progression of *d-sp* electron hybridization. For Fe on Li the spectrum presents a renormalized quasiparticle resonance near $E_F$ and lower Hubbard bands around -2 eV. In the Anderson model, the quasiparticle resonance peak near $E_F$ is identified with the Kondo effect, representing low-energy excitations that involve the spin degrees of freedom of the impurity and conduction electrons. Remarkably, compared to previous studies [10, 11, 17], this is the first time that both Hubbard and Kondo correlation fingerprints are observed in a metal impurity system, which we find reminiscent of a strongly correlated solid, e.g., $V_2O_3$ [18, 19]. Moreover, a most interesting and totally unique feature is found for the Fe-Na system: the photoemission spectrum in this case is a mixture of an atomic multiplet structure, quasiparticle bands near the Fermi level, and the low lying Hubbard bands around -2 eV. Such features, indicated by labels in Fig. 3 and later exemplified by our theoretical analysis, mark the transition between the localized and hybridized extremes of Fig. 1.

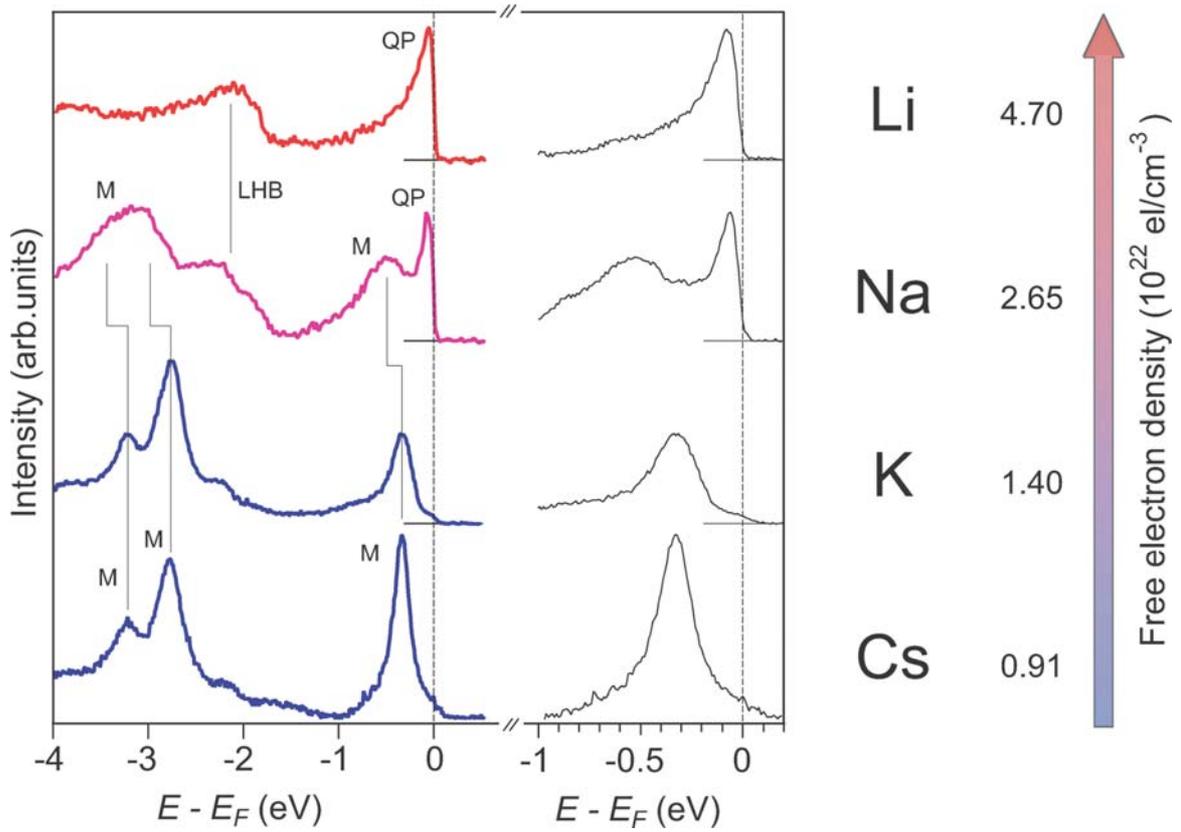

**Fig**. **3**. Host-dependent Fe impurity photoemission spectra. The data represent the impurity photoemission intensity after subtraction of the alkali background (see Fig. 2). The right scale reports the host free-electron density of Cs to Li (from Ref. [2]).

These data provide a model system to test general *ab-initio* theories of localized-to-itinerant electron behavior. In order to do so, one needs to incorporate both many-body effects from the strong local Coulomb interaction among *d*-electrons with hybridization effects due to the surrounding electronic bath of the host. All these many-body effects are beyond traditional electronic structure calculations; however, the recently developed first-principle Dynamical Mean

Field Theory (DMFT) [18-20] brings the prospect of investigating electronic correlations in realistic complex systems, studying the crossover from atomic-like behavior to band-like spectral properties. Progress using numerically exact many-body impurity solvers [15], which involve Quantum Monte Carlo or Exact Diagonalization methods, allows us to investigate such a transition in great detail.

To analyze the experimental results we have calculated the electronic structure of the Fe impurities in a bath of alkali electronic states using different impurity solvers of the DMFT method [18]. In this approach, the original many-body problem of the crystal is split into a *one-electron* problem for the lattice with a self-consistent local self-energy and a many-body *single-site* problem for a quantum impurity in an effective medium. We used the density functional method for estimating the electronic structure of different alkali crystals. A complete four-index intra-atomic Coulomb interaction matrix $U_{ijkl}$ is taken into account, with an average Hubbard parameter $U = 8$ eV and an exchange parameter $J = 0.85$ eV, which correspond to the optimal choice for the screened Coulomb interactions for Fe [18]. The corresponding impurity problem can be described by the following multi-orbital Anderson impurity model:

$$(1) \quad H_{imp} = \sum_{ij\sigma} \varepsilon_{ij}^d d_{i\sigma}^+ d_{j\sigma} + \frac{1}{2} \sum_{ijkl\sigma\sigma'} U_{ijkl} d_{i\sigma}^+ d_{j\sigma'}^+ d_{l\sigma'} d_{k\sigma} + \sum_{ik\sigma} \left( V_{ik} d_{i\sigma}^+ b_{k\sigma} + V_{ki} b_{k\sigma}^+ d_{i\sigma} \right) + \sum_{k\sigma} \varepsilon_k^b b_{k\sigma}^+ b_{k\sigma}$$

where $d^+$, $d$ and $b^+$, $b$ are creation and annihilation operators for correlated *d*- and bath *sp*-electrons, respectively, $\varepsilon^d$ and $\varepsilon^b$ are the one electron energies of correlated and bath electrons, and $V_{ik}$ is the hybridization matrix. Starting from density functional calculations of one-electron matrix elements of the effective Anderson Hamiltonian, we performed many-body calculations using different approximations. The most accurate scheme corresponds to the continuous time Quantum Monte Carlo solutions of the impurity Hamiltonian. In this method, the resulting Green function obtained in imaginary time should be analytically continued to the real energy axis, with the effect that all multiplet structures disappear. We also found that the simple Hubbard-I approximation [21] overestimates the tendency towards multiplet formation and hence it cannot explain the quasiparticle formation in the Fe-Li system. Finally, the most successful approach to describe the transition from atomic multiplet structures to a broad energy band is obtained with the Exact Diagonalization scheme for the solution of the impurity problem. We find that already a small number of orbitals in the fermionic bath is sufficient to change drastically the multiplet structure.

In Fig. 4 we show how the quasiparticle spectral function evolves as a function of the hybridization strength, V, using a single bath orbital. The negligibly weak hybridization (V~0 eV) reproduces the $d^7$ multiplet structure of the Fe-Cs and Fe-K systems with good accuracy. The small hybridization with the fermionic bath (V~0.2 eV) corresponds to the intermediate Fe-Na system with a combination of atomic multiplets and quasiparticle states at $E_F$, together with the new low Hubbard bands around -2 eV. Note that, according with the experiment, the Hubbard band and multiplet states broaden and shifts to higher binding energy as the spectral weight is redistributed towards the quasiparticle resonance at $E_F$. We speculate that a possible reason for the formation of a "mixed multiplet-band" feature is related to an anisotropic hybridization function. In theoretical calculations only a fully symmetric combination of Fe $d$-orbitals can hybridize with an $s$-like bath orbital. In the experimental situation such anisotropy can occur due to anisotropic hopping matrix elements between the Fe impurity and alkali metal surface atoms, which often have a complicated structure. Finally, the large hybridization with the fermionic bath (V~1 eV) corresponds to the Fe-Li system, where the atomic multiplet structure is removed and only the correlated quasiparticle and broad Hubbard bands around -2 eV are formed.

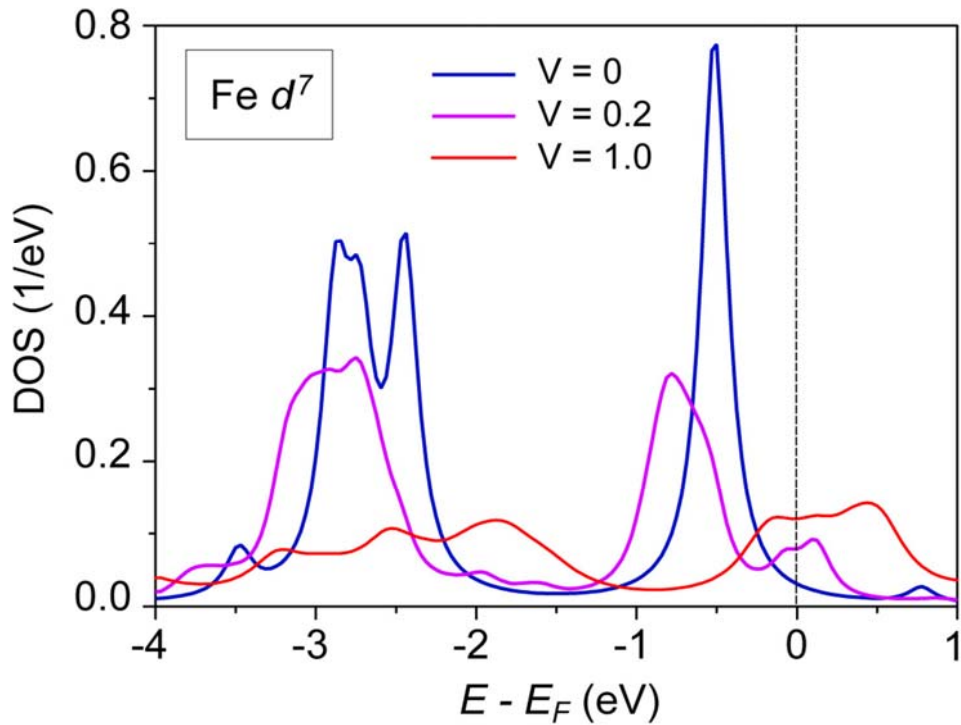

**Fig**. **4**. Theoretical spectral function of Fe impurities in alkali metal hosts. Colors refer to the electron conduction bath with zero (blue: V = 0 eV), small (purple: V = 0.2 eV) and large (red V = 1 eV) hybridization. The general Coulomb vertex is defined through the Slater parameters $F^0$ = 8 eV, $F^2$ = 7 eV, and $F^4$ = 5 eV, corresponding to $U$ = 8 eV and $J$ = 0.85 eV.

In conclusion, photoemission measurements of Fe impurities on alkali metal surfaces clearly show the effect of how atomic multiplets develop in solids when the relative strength of Coulomb repulsion and band formation effects are carefully balanced. The most interesting feature of such crossover behavior is related to the intermediate situation, where both a renormalized quasiparticle and a multiplet structure appear. We believe that this non-trivial situation is related with the anisotropy of the hybridization function to the fermionic bath and should be investigated in more detail in future theoretical research. The nice agreement between the experimental photoemission spectra and DMFT-like Exact Diagonalization impurity scheme proves the importance of considering full electron correlation effects as well as anomalies of the hybridization function in the interpretation of energy bands in transition-metal systems.

**ACKNOWLEDGEMNTS**

We acknowledge partial financial support through the EUROCORES SANMAG project of the European Science Foundation, SFB-668 of German Science Foundation and FOM (the Netherlands).